\documentclass[letterpaper, 10 pt,conference]{ieeeconf}
\IEEEoverridecommandlockouts
\usepackage[makeroom]{cancel}
\usepackage{amsmath,amssymb,amsfonts}
\usepackage{abraces}
\usepackage{algorithmic}
\usepackage{graphicx}
\usepackage{textcomp}
\let\labelindent\relax
\usepackage{enumitem}

\usepackage{xcolor}
\def\BibTeX{{\rm B\kern-.05em{\sc i\kern-.025em b}\kern-.08em
    T\kern-.1667em\lower.7ex\hbox{E}\kern-.125emX}}

\usepackage{stmaryrd} 

\newtheorem{theorem}{Theorem}
\newtheorem{lemma}{Lemma}

\newtheorem{assumption}{Assumption}

\usepackage{mathtools,bm}

\usepackage{tikz}
\usetikzlibrary{positioning, arrows.meta,calc}
\usepackage{pgfplots}

\usepackage{subfigure}

\usepackage[demo]{adjustbox}
\usepackage{float, caption, makecell, cellspace}
\usepackage{newtxtext, newtxmath} 
\usetikzlibrary{decorations.pathreplacing,angles,quotes}

\begin{document}
\setlist[itemize]{leftmargin=*}

\title{Directional excitability in Hilbert spaces}


\author{Gustave Bainier$^{1}$, Alessio Franci$^{1,2}$
\thanks{$^{1}$ Dept. of Electrical Engineering and Computer Science, University of Liège, 4000 Liège, Belgium; $^{2}$ WEL Research Institute, 1300 Wavre, Belgium, {\tt\small gustave.bainier@uliege.be}, {\tt\small afranci@uliege.be}
       }%
\thanks{This work was supported by the Belgian Government through the Federal Public Service Policy and Support.}
}

\maketitle

\begin{abstract} We introduce a generalized excitable system in which spikes can happen in a continuum of directions, therefore drastically enriching the expressivity and control capability of the spiking dynamics. In this generalized excitable system, spiking trajectories happen in a Hilbert space with an excitable resting state 
and spike responses that can be triggered in any direction as a function of the system's state and inputs. State-dependence of the spiking direction provide the system with a vanishing spiking memory trace, which enables robust tracking and integration of inputs in the spiking direction history. The model exhibits generalized forms of both Hodgkin's Type I and Type II excitability, capturing their usual bifurcation behaviors in an abstract setting.
In control engineering applications the proposed model facilitates both the sparseness of the actuation and its sensitivity to environmental inputs, as illustrate in a two-dimensional navigation task.
These results highlight the potential of the proposed generalized excitable model for excitable control in high- and infinite-dimensional spaces.
\end{abstract}


\section{Introduction}

Excitability is a fundamental property of biological neurons, allowing them to process and transmit information efficiently through spiking signals~\cite{kandel2000principles,Izhikevich2006}. The event-based nature of spiking signals is thought to be key for the speed and flexibility of biological nervous systems by enabling sparse and efficient communication, adaptive and on-demand computation, and fast and robust actuation~\cite{sterling2015principles}. Neuromorphic engineering aims at taking inspiration from the event-based, adaptive, and energy efficient nature of biological nervous system to design radically new kinds of sensors~\cite{Liu2010}, controllers~\cite{Sepulchre2022}, and computational principles~\cite{Schuman2022,Kudithipudi2025}. Like in their biological counterpart, the spikes of a neuromorphic controller enable fast, robust, and low-power decision-making in complex environments and constitute a step forward toward bio-inspired embodied intelligence~\cite{Bartolozzi2022}.

In virtually all spiking models, spikes can happen in only one direction, corresponding to a binary decision: to spike or not spike. However, when looking at the collective activity of neural populations, the population spike can happen in an extremely large number of directions, each associated with a different population-level spiking pattern, that encode, e.g., sensorimotor information~\cite{ayzenshtat2010precise,muller2018cortical} or perceptual decisions~\cite{bayones2024orthogonality}.

In order to increase the expressivity of simple excitable systems, this paper introduces a new, generalized model of excitable behavior, in which the spiking trajectories are occurring in a Hilbert space. In the absence of inputs, the excitable resting state is located at the origin, and a spike response can be initiated in any direction of the Hilbert space in which the model is defined. Spiking in a given direction can be considered as a fast, flexible and adaptive decision (to spike or not), over a continuum of options (the directions of the Hilbert space). The direction in which a spike is generated depends on two factors: the present state of the system, as determined by the past spiking history, and the inputs to the system. The vanishing memory provided by state-dependence allows the model to filter out input noise and jittering, ensuring robust, yet reactive, excitable response to inputs. 

To ensure analytical tractability of the proposed systems dynamics, directional spiking is obtained as the feedback interconnection of a fast and a slow subsystem:
\begin{itemize}
    \item A fast Hilbert-space dynamic for spike generation, realized as a gradient system whose potential determines the spiking threshold and the input integration dynamic.
    \item A slow one-dimensional dynamic, which integrates an increasing function of the norm of the Hilbert space state, and uses it to modulate the potential function of the Hilbert-space dynamic in order to terminate spikes and bring the Hilbert space state back to the excitable state.
\end{itemize}
These modeling choices are key for maintaining the model complexity low and to enable rigorous analysis of its behavior, without compromising the expressivity of the resulting spiking dynamics. In particular, the gradient system nature of the fast dynamic and the use of a one-dimensional slow dynamic make it possible to analyze the model behavior using a two-dimensional system, even when the spiking Hilbert space is infinite-dimensional.

To illustrate possible control applications of the proposed model, we test it as a directional spiking controller for a two-dimensional navigation task, in which case the spiking Hilbert space is the two-dimensional Euclidean space. This toy example provides a simple sensorimotor testbed to explore the performance of neuromorphic controllers before their applications to more complex robotics situations. Several neuromorphic methodologies are already being developed for navigation tasks~\cite{Yang2023,Novo2024}, and the controller proposed here is akin the one considered in~\cite{Cathcart2024,Giovanna}. The proposed model can be seen as a mid-point between the finite-dimensional, two-option (left or right) model used in~\cite{Cathcart2024} and the infinite-dimensional, infinite option (all possible planar directions) model used in~\cite{Giovanna}. The proposed model is finite-dimensional, yet capable of making spiking decision in any direction of the plane.


This paper is organized as follows. Section~\ref{sec:sysdef} introduces the generalized excitable system. 
Section~\ref{sec:excitability} studies the system's excitability and illustrates how it can exhibit both Hodgkin's Type I and Type II excitability.
Section~\ref{sec:2dnav} uses the proposed excitable system for a two-dimensional navigation task. Finally, Section~\ref{sec:concl} concludes the paper.

\section{System definition and input-output characterization} \label{sec:sysdef}

    \newsavebox{\genericfilt}
\savebox{\genericfilt}{%
    \begin{tikzpicture}
        \draw[domain=0:0.65, smooth, variable=\x, very thick, red, densely dotted] 
            plot ({-0.3+5*\x*\x-6*\x*\x*\x*\x}, {\x});
        \draw[domain=0.65:0.98, smooth, variable=\x, very thick, red] 
            plot ({-0.3+5*\x*\x-6*\x*\x*\x*\x}, {\x});
        \draw[very thick, red] (-0.3,0)--(0.9,0) ;
        \draw[very thick, densely dotted, red] (-0.3,0)--(-1,0) ;
            \draw[->] (-1,0)--(1,0) node[above left]{$x_s$};
            \draw[->] (0,0)--(0,1.2) node[ right]{$\lVert \mathbf{x} \rVert$};
    \end{tikzpicture}%
}

\begin{figure}[t]
    \begin{center}
    \begin{tikzpicture}[auto, node distance=2cm, >=Latex]
        \tikzstyle{xy} = [coordinate]
        \tikzstyle{Gblock} = [rectangle, draw,  fill=red!15, rounded corners, minimum height=3em, minimum width=5em]
        \tikzstyle{block} = [rectangle, draw, fill=blue!15, 
            text width=3em, text centered, rounded corners, minimum height=3em, minimum width=3em]
        \tikzstyle{sum} = [draw, circle, node distance=1cm]
        \node [Gblock, name=plant] {\usebox{\genericfilt}};
        \node [block, name=LPF,  below left= 0.5cm and -1cm of plant] {$\frac{1}{\tau_s s+1}$};
        \node [block, name=g,  below right= 0.5cm and -1cm of plant] {$g(\lVert \cdot{} \rVert )$};
        \node [xy, left=1.2cm of plant.180, name=input] {};
        \node [xy, below right=1.3cm and 0.5cm of input, name=input2] {};
        \node [xy, right=1.2cm of plant, name=output] {};
        \node [xy, left=0.4cm of output, name=output2] {};

        \draw [->] (input) -- node [name=plant_input] {$\mathbf{u}(t)$} (plant);
        \draw [-] (plant) --  node [name=plant_output] {$\mathbf{x}(t)$} (output2);
        \draw [->] (output2) -- (output);
        \draw [->] (output2) |- (g);
        \draw [->] (g.180) |- (LPF);
        \draw [-] (LPF.180) -| node [name=xs, yshift=0.9cm, xshift=0.7cm] {$x_s(t)$}  (input2);
        \draw [->] (input2) -|  (plant.270);

        \draw[decoration={brace,mirror,raise=5pt},decorate, thick] (2.3,-1) -- node[right=6pt] {\eqref{eq:sys1}} (2.3,1);
        \draw[decoration={brace,mirror,raise=5pt},decorate, thick] (2.3,-2.6) -- node[right=6pt] {\eqref{eq:sys2}} (2.3,-1.1);
    \end{tikzpicture}
\end{center}
    \caption{ \small Block diagram of the system \eqref{eq:sys}. The red block at the top stands for gradient system \eqref{eq:sys1} and displays the bifurcation diagram of \eqref{eq:potentialDyn} with respect to $x_s$, where the full (resp. dotted) red lines are a continuum of asymptotically stable (resp. unstable) equilibria. The two blue blocks at the bottom represent the slow dynamic \eqref{eq:sys2} modulating \eqref{eq:sys1}.}
    \label{fig:bloc}
\end{figure}
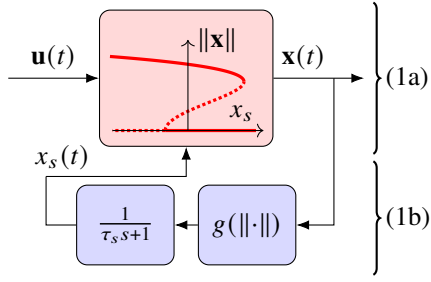

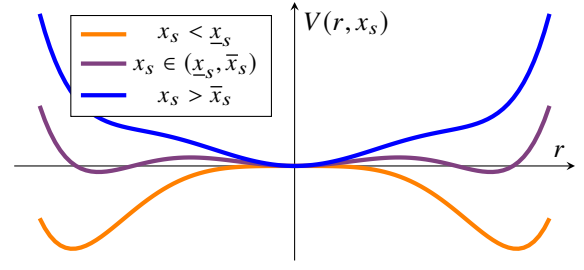
\begin{figure}[t!]
    \centering
    \begin{tikzpicture}[scale=1]
    \begin{axis}[
        xlabel={$r$}, ylabel={$V(r,x_s)$},
        axis lines=middle,
        width=9cm, height=5cm,
        legend style={at={(0.1,0.95)}, anchor=north west},
        grid=none,
        samples=100,
        domain=-1.3:1.3,
        enlarge x limits=0.05,
        enlarge y limits=0.05, 
        ticks=none,
        legend style={font=\small}
    ]
    \foreach \a/\c in {-1/0, 0.1/50, 1/100} {
    \edef\temp{
        \noexpand
        \addplot[ultra thick,color=blue!\c!orange] {(1+\a)*x^2 - 2.5*x^4 + 1.3*x^6};}
    \temp
    }
    \addlegendentry{$x_s < \underline{x}_s$};
    \addlegendentry{$x_s \in (\underline{x}_s,\overline{x}_s)$};
    \addlegendentry{$x_s > \overline{x}_s$};
    \end{axis}
    \end{tikzpicture}
    \caption{ \small Potential $V$ of the gradient system \eqref{eq:potentialDyn} at three $x_s$ values.}
    \label{fig:potential} 
\end{figure}

Consider the following dynamics.
\begin{subequations} \label{eq:sys}
    \begin{align}    
        \tau\dot{\mathbf{x}} &= - \frac{\partial}{\partial \mathbf{x}} V\left(\lVert \mathbf{x}\rVert,x_s\right)+\alpha\mathbf{u}  \label{eq:sys1}\,, \\
        \tau_s \dot{x}_s &=- x_s+g \left(\lVert \mathbf{x}\rVert \right)\,. \label{eq:sys2}
    \end{align}
\end{subequations}
\begin{itemize}
    \item The state $\mathbf{x}(t)$ and the input $\mathbf{u}(t)$ at each instant $t$ belong to a $\mathbb{R}$-Hilbert space $\left(\mathcal{H}, \langle \cdot{}, \cdot{} \rangle \right)$, with $\lVert \cdot{}\rVert$ the norm induced by the scalar product $\langle \cdot{}, \cdot{} \rangle$.
    \item The potential $V\in \mathcal{C}^1(\mathbb{R}^2,\mathbb{R})$ is chosen so that Assumptions~\ref{ass:V1} and~\ref{ass:V2} below are satisfied.
    \item The variable $x_s(t) \in \mathbb{R}$ of \eqref{eq:sys2} serves as modulating parameter for the shape of the potential $V$ of the gradient system~\eqref{eq:sys1} (Figure~\ref{fig:potential}).
    \item The time constants $(\tau,\tau_s)$ adjust the timescale separation of the system, with $\tau_s\gg \tau>0$, so that $x_s$ evolves more slowly than ${\bf x}$, and $\alpha$ is the input gain.
    \item The function $g\in \mathcal{C}^0(\mathbb{R}_{\geq 0},\mathbb{R}_{\geq 0})$ is chosen to be monotone increasing, with $g(0)=0$.
\end{itemize}

System~\eqref{eq:sys} consists of a gradient system \eqref{eq:sys1} modulated by the slow one-dimensional dynamic \eqref{eq:sys2}. It is assumed to possess a solution on the maximum interval of existence $\mathbb{R}_{\geq 0}$ for all initial condition $(\mathbf{x}_0,x_{s,0})\in \mathcal{H}\times \mathbb{R}_{\geq 0}$. This can be demonstrated using the standard assumptions of the Cauchy-Lipschitz theorem \cite{Feng2016}. The block diagram of \eqref{eq:sys} is presented in Figure~\ref{fig:bloc}.

Two key assumptions are imposed on the system. The first one is purely technical and serves only in the proof of Lemma~\ref{lem:technical}. The second one is a necessary (though not sufficient) condition for the system to be excitable.

\begin{assumption} \label{ass:V1}
    There exists two constants $(a,b) \in\mathbb{R}_{>0} \times \mathbb{R}_{\geq 0}$ and an increasing function $h \in \mathcal{C}^0(\mathbb{R}_{\geq 0},\mathbb{R}_{\geq 0})$ with $h(0)=0$, such that for all $x_s \in \mathbb{R}_{\geq 0}$:
\begin{equation}
    ar-b \leq \frac{\partial}{\partial r} V(r,x_s) \label{eq:ineq_ass1} 
\end{equation}
and for all $\overline{r}\in \mathbb{R}_{\geq 0}$, $r\in[0,\overline{r}]$, and $x_s \in \left[0,\lVert x_{s,0}\rVert +g\left(\overline{r}\right)\right]$:
\begin{equation}
       \frac{\partial}{\partial r} V(r,x_s) \leq  h(\overline{r})r \label{eq:ineq_ass2}
\end{equation}
\end{assumption}

\begin{assumption} \label{ass:V2}
The one-dimensional gradient system
    \begin{equation}\label{eq:potentialDyn}
        \dot{r}=-\frac{\partial}{\partial r}V(r,x_s)
    \end{equation}
has an equilibrium at $r=0$ that undergoes a subcritical pitchfork bifurcation with respect to its parameter $x_s$ at $x_s=\underline{x}_s$. In particular, there exists $\overline x_s > \underline x_s$ such that:
\begin{itemize}
    \item If $x_s<\underline x_s$, then~\eqref{eq:potentialDyn} has two locally asymptotically stable equilibria at $r=\pm r^*$.
    \item If $\underline x_s< x_s < \overline x_s$, then~\eqref{eq:potentialDyn} has three locally asymptotically stable equilibria at $r=-r^*,0,r^*$. Furthermore, $\frac{\partial r^*}{\partial x_s}<0$.
    \item If $x_s>\overline x_s$, then $r=0$ is a globally asymptotically stable equilibrium of~\eqref{eq:potentialDyn}.
\end{itemize}
\end{assumption}

We have the following two theorems that characterize the behavior of~\eqref{eq:potentialDyn}. Their proof can be found in the Appendix.

\begin{theorem}[Spiking norm dynamics] \label{lem:normSpike}
    The dynamic of the norm of $\mathbf{x}$ in~\eqref{eq:sys} reads, for $\mathbf{x} \neq 0$,
\begin{subequations} \label{eq:2dreduced_dyn}
\begin{align}    
    \tau\dot{\aoverbrace[L1R]{\lVert \mathbf{x}\rVert}}  &=- \frac{\partial}{\partial \lVert\mathbf{x} \rVert} V\left(\lVert \mathbf{x}\rVert,x_s\right)+\alpha\tilde{u} \label{eq:2dreduced_dyn1}\\
    \tau_s \dot{x}_s &=- x_s+ g\left(\lVert \mathbf{x}\rVert\right) \label{eq:2dreduced_dyn2}
\end{align}
\end{subequations}
where $\tilde{u}(t) \in \mathbb{R}$ is an input signal defined by $\tilde{u} \triangleq \cos(\mathbf{x},\mathbf{u})\lVert \mathbf{u} \rVert$.
\end{theorem}

\begin{lemma}\label{lem:technical} Let $x_{s,0} \in \mathbb{R}_{\geq 0}$ and $\mathbf{u} \in \mathcal{H}$ taken constant such that $\lVert \mathbf{u} \rVert >b/\alpha$, then there exists a time $t \geq 0$ above which there exists $\underline{r},\overline{r}\in\mathbb{R}_{>0}$ such that $\lVert \mathbf{x} \Vert \in [\underline{r}, \overline{r}]$.
\end{lemma}

\begin{theorem}[Direction dynamics] \label{lem:dirSpike} 
    The dynamic of the cosine between $\mathbf{x}$ and a constant input $\mathbf{u} \in \mathcal{H}$ in \eqref{eq:sys} reads, for $\mathbf{x} \neq 0$,
    \begin{equation} \label{eq:cosdyn}
        \frac{\tau}{\alpha}\dot{\aoverbrace[L1R]{\cos(\mathbf{x},\mathbf{u})}} = \frac{\lVert \mathbf{u} \rVert}{ \lVert \mathbf{x} \rVert} \left(1-\cos^2(\mathbf{x},\mathbf{u}) \right)
    \end{equation}
    which is a Riccati differential equation whose unique solution is given by
    \begin{equation} \label{eq:soleqdiffcos}
        \cos(\mathbf{x},\mathbf{u}) = \tanh \left(\frac{\alpha \lVert \mathbf{u}\rVert}{\tau}\int_0^t \lVert \mathbf{x}(s) \Vert^{-1}ds+c_0\right)
    \end{equation}
    with $c_0 =  \tanh ^{-1}(\cos(\mathbf{x}_0,\mathbf{u}))$ \cite{Zaitsev2002}. In particular, if the conditions of Lemma~\ref{lem:technical} are satisfied, then $\lim_{t\to +\infty} \cos(\mathbf{x},\mathbf{u}) = 1$.
\end{theorem}

\begin{figure}[t]
    \centering
    \includegraphics[width=\linewidth]{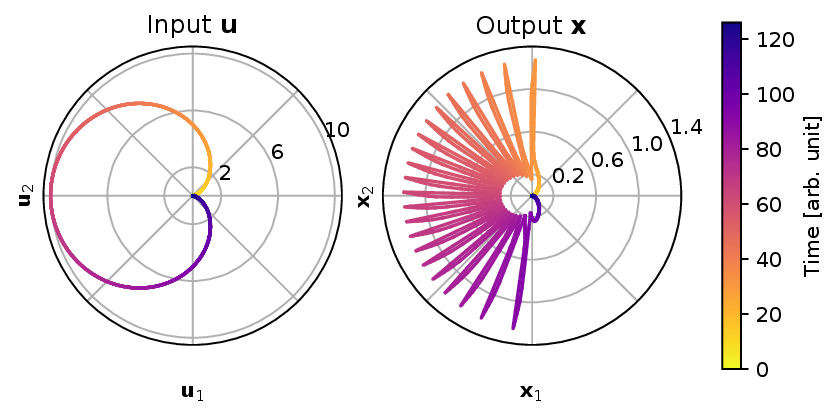}
    \caption{ \small Trajectories of the input $\mathbf{u}$ and output $\mathbf{x}$ of \eqref{eq:sys} for $\mathcal{H} = \mathbb{R}^2$, with the input $\mathbf{u}=(1-\cos\theta_u)\left(\cos\theta_u , \sin\theta_u\right)$ where $\theta_u = t/20$, and $(\mathbf{x}_0,x_{s,0}) = (0,0,1)$. The system is parameterized by $(V,g)=\eqref{eq:Vg}, \tau = 0.05, \tau_s = 2, \alpha = 1/4, \beta_1 = 3, \beta_2 = 1.5, \beta_3 = 1.5$ and exhibits Type I excitability.}
    \label{fig:IO}
\end{figure}

Theorem~\ref{lem:normSpike} shows that the dynamic of the norm of $\mathbf{x}$ reduces to a simple two-dimensional system that can exhibit spiking behavior for suitable choices of the functions $(V,g)$~\cite{Izhikevich2006}. Theorem~\ref{lem:dirSpike} shows that the spiking state of the system $\mathbf{x}$ asymptotically aligns with the direction of its input.
Taken together, these two results show that for a sufficiently large input strength, the generalized excitable system~\eqref{eq:sys} is capable of producing trains of directional spikes in a direction that asymptotically approaches that of the input. This property is illustrated in Figure~\ref{fig:IO} for $\mathcal{H}=\mathbb{R}^2$, taken with its usual scalar product.


\section{Type I and II excitability} \label{sec:excitability}

This section shows that the generalized excitable system~\eqref{eq:sys} can exhibit both Hodgkin's Type I and Type II excitability~\cite{Hodgkin1948}. This is accomplished through a standard phase-plane analysis of the two-dimensional dynamics \eqref{eq:2dreduced_dyn} under a constant input signal $\tilde{u}$, following the typical approach used for reduced models of neuronal excitability \cite{Izhikevich2006}. For the rest of the paper, the following functions
\begin{subequations} \label{eq:Vg}
    \begin{align}
        V(r,x_s) &= \frac{1}{2}(1+x_s)r^2-\frac{1}{4}\beta_1 r^4+\frac{1}{6}\beta_2 r^6 \\
        g(r) &= \beta_3 r^4
    \end{align}
\end{subequations}
are considered in~\eqref{eq:sys}, but are by no means the only possible choices to obtain an excitable behavior.
The choice of $\{\beta_i\}_{1 \leq i \leq 3}$ determines the excitability properties of the system, including their Hodgkin's excitability Type. Using~\eqref{eq:Vg}, the norm-spiking dynamics~\eqref{eq:2dreduced_dyn} of~\eqref{eq:sys} reads
\begin{subequations} \label{eq:2dreduced_dyn_ex}
\begin{align}    
    \tau \dot{\aoverbrace[L1R]{\lVert \mathbf{x}\rVert}}  &=-(1+x_s)\lVert \mathbf{x}\rVert+\beta_1\lVert \mathbf{x}\rVert^3 - \beta_2 \lVert \mathbf{x}\rVert^5+\alpha\tilde{u} \label{eq:2dreduced_dyn1_ex}\\
    \tau_s \dot{x}_s &=- x_s+ \beta_3 \lVert \mathbf{x}\rVert^4\label{eq:2dreduced_dyn2_ex}
\end{align}
\end{subequations}
which enables a straightforward phase-plane analysis. Note that:
\begin{itemize}
    \item The $\lVert \mathbf{x} \rVert$-nullcline is the bifurcation diagram of \eqref{eq:2dreduced_dyn1} (here \eqref{eq:2dreduced_dyn1_ex}) with respect to $x_s$, as portrayed for $\mathbf{u}=0$ in the red block at the top of Figure~\ref{fig:bloc}.
    \item The $x_s$-nullcline is the plot of the $g$ function $x_s= g(\lVert \mathbf {x}\rVert)$.
\end{itemize}
Hence, both nullclines can be shaped precisely using rigorous analytical and bifurcation-theoretical arguments, if needed.

\subsection{Type I}

\begin{figure}[t!]
    \centering
    \includegraphics[width=0.9\linewidth]{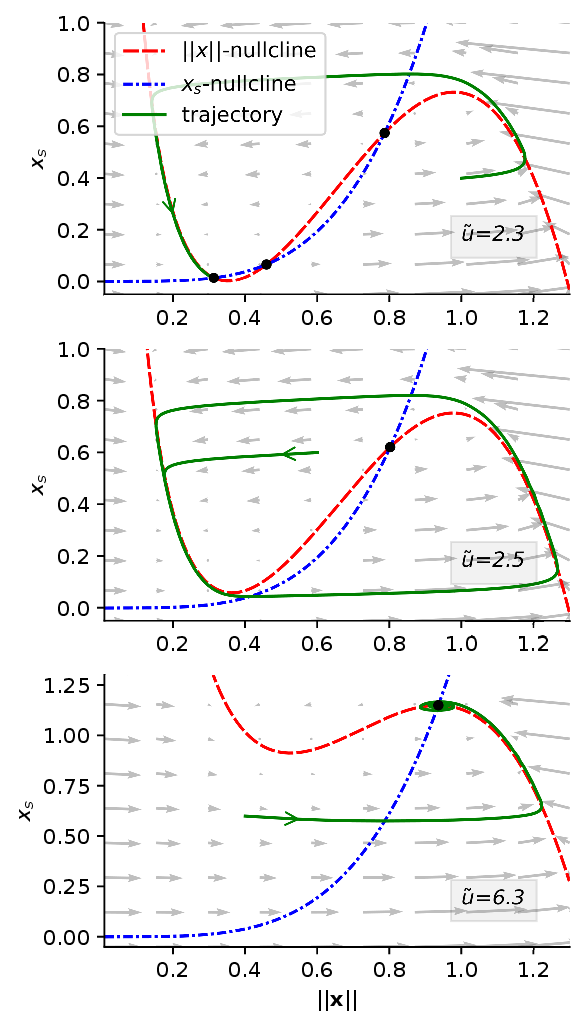}
    \caption{ \small Phase-planes of a system \eqref{eq:2dreduced_dyn} exhibiting Type I excitability, and considered at three distinct input values $\tilde{u}$, resulting in three possible qualitatively distinct behaviors. The functions $(V,g)$ are taken as \eqref{eq:Vg} and the system is parameterized by $\tau = 0.1, \tau_s = 3, \alpha = 0.1, \beta_1 = 3, \beta_2 = 1.5, \beta_3 = 1.5$.}
    \label{fig:type1}
\end{figure}

Type I excitability is characterized by a continuous frequency response to a constant input $\tilde{u}$, meaning the system \eqref{eq:2dreduced_dyn_ex} spikes at an arbitrarily low frequency $f>0$ when the input $\tilde{u}$ is just above a certain threshold, the rheobase $\underline{\tilde{u}}$. This behavior arises from a saddle-node bifurcation on an invariant cycle, allowing smooth transitions in the firing rate. This excitability type is well-suited for encoding gradual changes in the input, as illustrated by Figure~\ref{fig:IO}. As the input slowly ramps up, the spiking frequency rises, while the spikes' width decreases, until a Hopf bifurcation occurs at a new threshold $\overline{\tilde{u}}$, beyond which the spiking stops. The following qualitative behaviors can be deduced from the phase-plane analysis of the system:
\begin{itemize}
    \item For a constant input $\tilde{u}\in [0,\underline{\tilde{u}})$, the system acts as a simple integrator of the input near its resting equilibrium. A single spike can still be induced by a sufficiently strong transient input. The phase-plane of this behavior is portrayed at the top of Figure~\ref{fig:type1}.
    \item For a constant input $\tilde{u}\in (\underline{\tilde{u}},\overline{\tilde{u}})$, the system exhibits a tonic spiking activity characterized by an asymptotically stable limit cycle whose frequency grows with the input strength. The phase-plane of this behavior is portrayed at the middle of Figure~\ref{fig:type1}.
    \item For a constant input $\tilde{u}\in (\overline{\tilde{u}},+\infty)$, the system exhibits damped oscillations around the new globally asymptotically stable equilibrium. The phase-plane of this behavior is portrayed at the bottom of Figure~\ref{fig:type1}.
\end{itemize} 
Here, Type I excitability is obtained with $\beta_1=3, \beta_2=1.5, \beta_3=1.5$. Using the nullcline analysis of~\eqref{eq:2dreduced_dyn_ex}, one can estimate that:
\begin{equation}
    \underline{\tilde{u}}\approx\frac{1}{\alpha}0.24\dots \qquad \overline{\tilde{u}}\approx\frac{1}{\alpha} 0.62\dots
\end{equation} 

\subsection{Type II}

\begin{figure}[t!]
    \centering
    \includegraphics[width=0.9\linewidth]{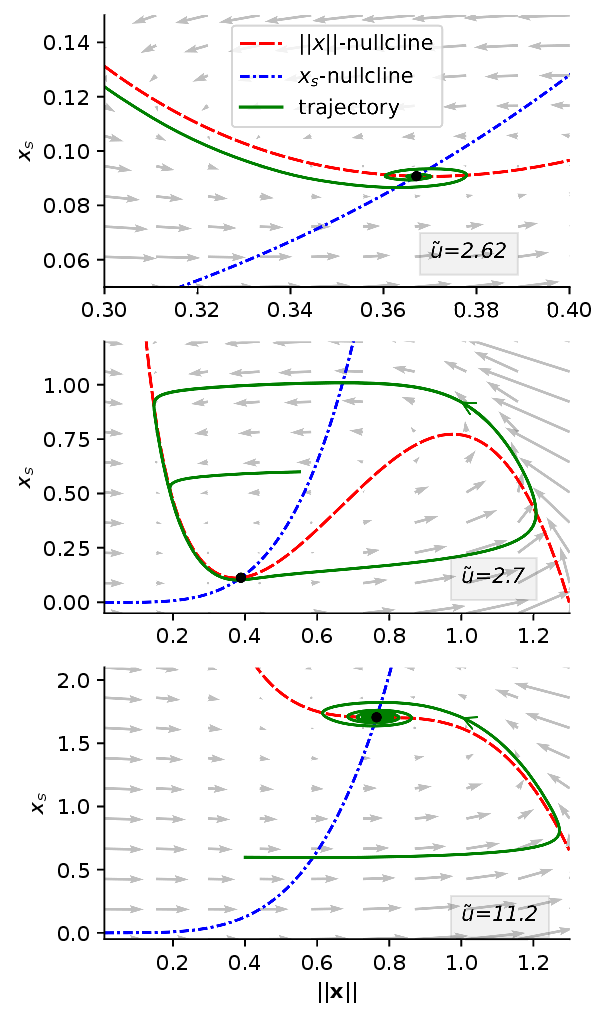}
    \caption{ \small Phase-planes of a system \eqref{eq:2dreduced_dyn} exhibiting Type II excitability, and considered at three distinct input values $\tilde{u}$, resulting in three possible qualitatively distinct behaviors. The functions $(V,g)$ are taken as \eqref{eq:Vg} and the system is parameterized by $\tau = 0.1, \tau_s = 3, \alpha = 0.1, \beta_1 = 3, \beta_2 = 1.5, \beta_3 = 5$.}
    \label{fig:type2}
\end{figure}

Type II excitability exhibits a discontinuous frequency response to constant inputs $\tilde{u}$ as the input strength is increased. System~\eqref{eq:2dreduced_dyn_ex} spikes at a finite frequency $f \geq f_0 > 0$ as soon as the input $\tilde{u}$ is above the rheobase $\underline{\tilde{u}}$. In this subsection, we made the Type II behavior result from a Hopf bifurcation, which makes the response of the system more resonant as compared to the saddle-node bifurcation of a Type I system. Type II excitability is better suited for synchronization and rhythmic network activity rather than fine-tuned signal encoding. Nonetheless, as the input slowly ramps up, the spiking frequency still rises, while the spikes' width decreases, until another Hopf bifurcation occurs at a new threshold $\overline{\tilde{u}}$, beyond which the spiking stops. The following qualitative behaviors can be deduced from the phase-plane analysis of the system:
\begin{itemize}
    \item For a constant input $\tilde{u}\in [0,\underline{\tilde{u}})$, the system is globally asymptotically stable and acts as a simple resonator of the input near its resting equilibrium. A single spike can still be induced by a transient input which is either sufficiently large, or resonating with the system. The phase-plane of the near-to-equilibrium oscillations is portrayed at the top of Figure~\ref{fig:type2}.
    \item The cases $\tilde{u}\in (\underline{\tilde{u}},\overline{\tilde{u}})$ and $\tilde{u}\in (\overline{\tilde{u}},+\infty)$ are qualitatively similar to their Type I counterparts. The phase-planes of these behaviors are respectively portrayed at the middle and at the bottom of Figure~\ref{fig:type2}.
\end{itemize} 
This Type II excitability is here obtained with $\beta_1=3, \beta_2=1.5,  \beta_3=5$. Using the nullcline analysis of \eqref{eq:2dreduced_dyn_ex}, one can estimate that:
\begin{equation}
\underline{\tilde{u}} \approx \frac{1}{\alpha}0.26\dots \qquad \overline{\tilde{u}} \approx\frac{1}{\alpha}1.11\dots
\end{equation} 

\section{Application to $2d$ navigation with sparse actuation} \label{sec:2dnav}

\begin{figure}[t]
    \centering
    \includegraphics[width=\linewidth]{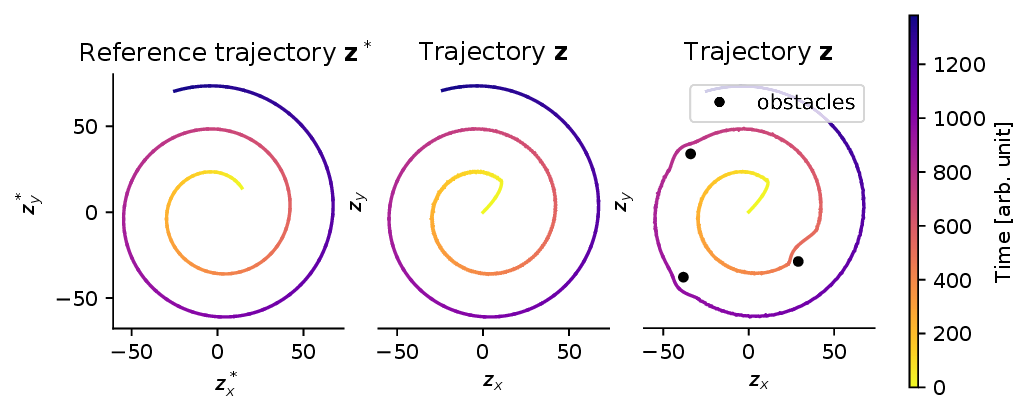}
    \caption{ \small  Two-dimensional view of the mobile robot trajectory $\mathbf{z}=(z_x,z_y)$ on the plane $\mathbb{R}^2$ with and without obstacles (resp. right and middle), compared to a reference trajectory $\mathbf{z}^*=(z_x^*,z_y^*)$ (left).}
    \label{fig:traj}
\end{figure}

This section considers a mobile robot on a two-dimensional plane. The robot's position is given by the vector $\mathbf{z} = (z_x,z_y)\in \mathbb{R}^2$. The robot's acceleration, $\ddot{\mathbf{z}}$, is assumed to be controlled by a two-dimensional ``all-or-none'' actuator $\mbox{act}(\mathbf{u})\in\mathbb{R}^2$ that is either zero or on the unit circle of $\mathbb{R}^2$. Using the excitability mechanism of this paper, a controller is obtained for ${\bf u}$ that ensures both:
\begin{itemize}
    \item sparse actuation in time, so that the robot only controls its acceleration when necessary;
    \item sensitivity to environmental inputs, so that the robot can dynamically adapt its trajectory to changing environments and, in particular, to the presence of obstacles.
\end{itemize}
The controlled robot dynamics are
\begin{equation} \label{eq:actuation}
    \ddot{\mathbf{z}} = -\gamma \dot{\mathbf{z}}+\alpha_{act} \,\mbox{act}(\mathbf{u}), \quad \mbox{where } \mbox{act}(\mathbf{u})\triangleq 1_{\lVert \mathbf{u} \rVert \geq S} \frac{\mathbf{u}}{\lVert \mathbf{u} \rVert}
\end{equation}
with $1_{\lVert \mathbf{u} \rVert \geq S} \triangleq 1$ if $\lVert \mathbf{u} \rVert \geq S$ and $ 1_{\lVert \mathbf{u} \rVert \geq S}=0$ if $\lVert \mathbf{u} \rVert < S$. To foster sparse actuation and good sensitivity to environmental inputs, the controller dynamic is assumed to follow the generalized excitability model \eqref{eq:sys}, as follows:
\begin{subequations} \label{eq:ctrl}
    \begin{align}    
        \tau\dot{\mathbf{u}} &= - \frac{\partial}{\partial \mathbf{u}} V\left(\lVert \mathbf{u}\rVert,u_s\right)+\mathbf{v} \\
        \tau_s \dot{u}_s &=- u_s+g \left(\lVert \mathbf{u}\rVert \right)
    \end{align}
\end{subequations}
The parameters of the excitable controller~\eqref{eq:ctrl} are chosen so that it exhibits Type I excitability, which promotes lower spiking frequency and therefore sparser actuation.
The input $\mathbf{v}$ to the excitable controller~\eqref{eq:ctrl} is defined depending on the robot's task.
 
\subsection{Trajectory tracking with sparse actuation}

\begin{figure}[t]
    \centering
    \includegraphics[width=\linewidth]{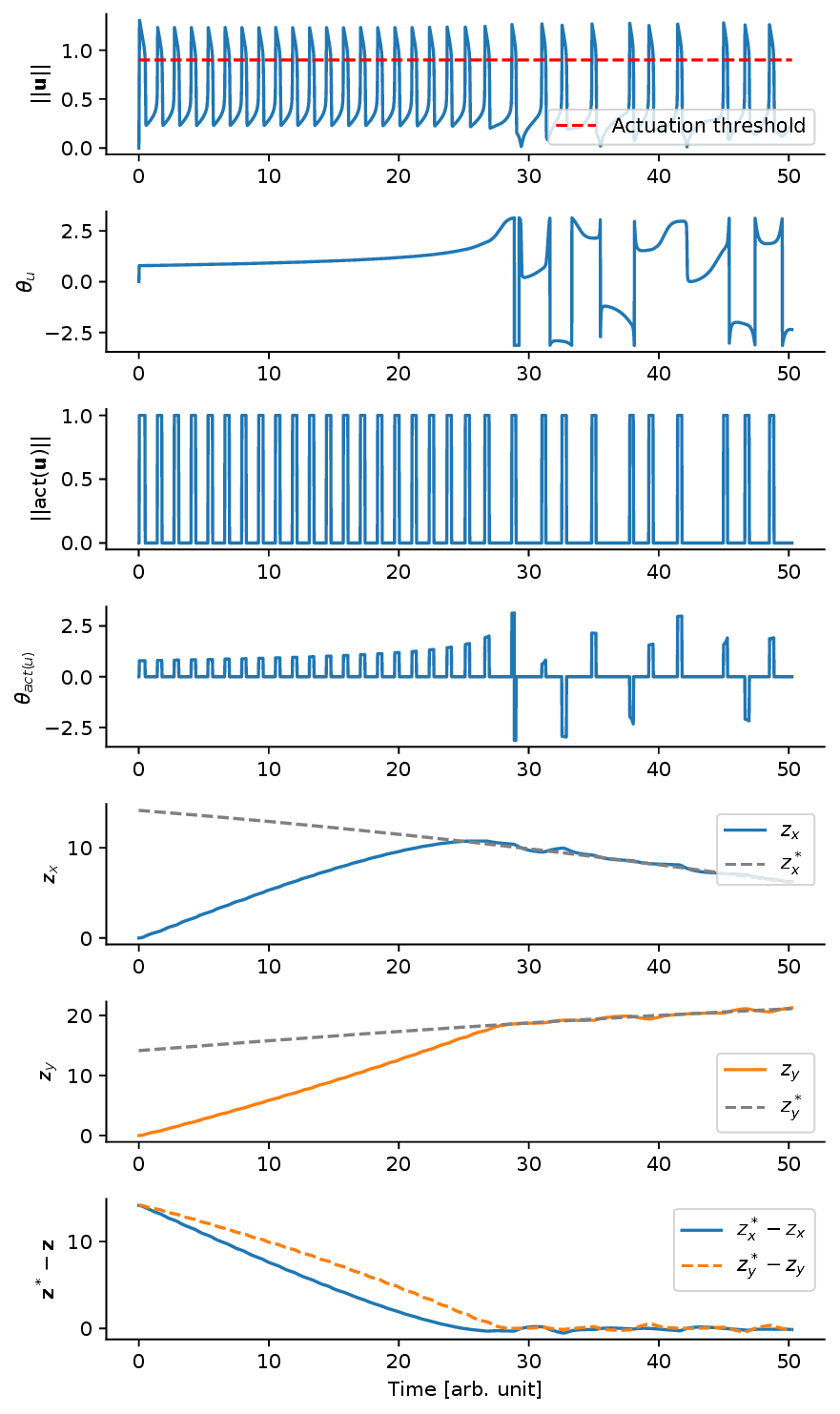}
    \caption{ \small  From top to bottom, time evolution of: the norm $\lVert \mathbf{u} \rVert$ and polar angle $\theta_u$ of the state of the excitable controller \eqref{eq:ctrl}, the norm $\lVert\mbox{act}( \mathbf{u})\rVert$ and polar angle $\theta_{\mbox{act}(u)}$ of the actuator defined in \eqref{eq:actuation}, a comparison between the robot's position $\mathbf{z}$ and target's position $\mathbf{z}^*$ respectively in $x$ and $y$ coordinates, and finally both coordinates of the position error $\mathbf{z}^*-\mathbf{z}$.}
    \label{fig:var1}
\end{figure}

The goal of this section is to design a controller input ${\bf v}$ such that the robot can follow a reference trajectory $\mathbf{z}^*$ on the plane.
A possible solution is to define
\begin{equation}
   \mathbf{v} =  k_1 \frac{\mathbf{z}^*-\mathbf{z}}{\varepsilon+ \|\mathbf{z}^*-\mathbf{z}\|}
\end{equation}
In other words, the excitable model~\eqref{eq:ctrl} receives as input the instantaneous direction of its target trajectory multiplied by the gain $k_1$. The regularization constant $\varepsilon>0$ appearing at the denominator eliminates the singularity at $\mathbf{z}=\mathbf{z}^*$. The functions $(V,g)$ are taken as in~\eqref{eq:Vg} and the system is parameterized by $\tau = 0.01, \tau_s = 1, \beta_1 = 3, \beta_2 = 1.5, \beta_3 = 1.5, k_1 = 0.4, \gamma = 1, \alpha_{act}  = 3, S=0.9, \varepsilon = 0.1$. The reference trajectory is defined by $\mathbf{z}^*(t) = (20+t/25) (\cos \theta^*(t),\sin \theta^*(t))$ with $\theta^*(t) = t/100+\pi/4$, and the mobile robot and its controller are initialized with $(\mathbf{z}_0,\dot{\mathbf{z}}_0, \mathbf{u}_0,u_{s,0}) = 0$. The trajectory $\mathbf{z}(t)$ of the mobile robot is compared to the reference trajectory $\mathbf{z}^*(t)$ in Figure~\ref{fig:traj}. Both the tracking performance and the sparseness of the actuation can be appreciated in Figure~\ref{fig:var1}.

\subsection{Responsiveness to a changing environment}

When the mobile robot's task is to follow a reference trajectory $\mathbf{z}^*$ while also avoiding a set of $n_o$ obstacles $\{\mathbf{z}^{o_{i}}\}_{0\leq i \leq n_o}$, the excitable controller input can be modified as follows  
\begin{equation}
  \mathbf{v} = k_1 \frac{\mathbf{z}^*-\mathbf{z}}{\varepsilon+ \|\mathbf{z}^*-\mathbf{z}\|}+k_2\sum_{i =1}^{n_o} \frac{\mathbf{z} -\mathbf{z}^{o_{i}}}{\|\mathbf{z} -\mathbf{z}^{o_{i}}\|^2}
\end{equation}
The closed-loop system is simulated with the same parameters as in the previous subsection, with $n_o = 3$ and $k_2 = 1.3$. The trajectory $\mathbf{z}$ of the mobile robot is compared to the reference trajectory $\mathbf{z}^*$ in Figure~\ref{fig:traj}, where we can observe that the three obstacles are correctly avoided.

\section{Conclusions and Perspectives}\label{sec:concl}

This paper introduced a novel generalization of excitable systems in a Hilbert space. The model features an excitable resting state close to the origin, with spike responses triggered in any direction based on the system's state and inputs. This system can exhibit both Hodgkin's Type I and Type II excitability, by capturing their characteristic bifurcation behaviors in a reduced two-dimensional model. Moreover, the model's applicability was demonstrated in a two-dimensional navigation task, where it enabled both sparse actuation and sensitivity to environmental inputs. These findings underscore the potential of this framework for neuro-inspired control schemes, particularly in high- and infinite-dimensional spaces. These more advanced applications remain to be investigated in future works.

\bibliographystyle{ieeetr}
\bibliography{refs}

  \appendix
  \subsection{Proof of Theorem 1}
For $\mathbf{x} \neq 0$, the following equalities hold:
\begin{equation}
    \tau \dot{\aoverbrace[L1R]{\lVert \mathbf{x}\rVert}} = \frac{\langle \dot{\mathbf{x}},\mathbf{x} \rangle}{\lVert \mathbf{x}\rVert}= -\frac{1}{\lVert \mathbf{x}\rVert} \left\langle  \frac{\partial}{\partial \mathbf{x}} V\left(\lVert \mathbf{x}\rVert,x_s\right),\mathbf{x} \right\rangle +\alpha\frac{\langle \mathbf{u},\mathbf{x} \rangle}{\lVert \mathbf{x}\rVert}
\end{equation}
by the chain rule:
\begin{equation} \label{eq:chainrule} 
         \frac{\partial}{\partial \mathbf{x}} V\left(\lVert \mathbf{x}\rVert,x_s\right) = \frac{\partial }{\partial\lVert \mathbf{x}\rVert} V\left(\lVert \mathbf{x}\rVert,x_s\right)\frac{\mathbf{x}}{\lVert \mathbf{x}\rVert}
\end{equation}
and by definition, $\cos(\mathbf{x},\mathbf{u}) \triangleq \frac{\langle \mathbf{u},\mathbf{x} \rangle}{\lVert \mathbf{u}\rVert\lVert \mathbf{x}\rVert}$. This finally provides:
\begin{equation}
 \tau \dot{\aoverbrace[L1R]{\lVert \mathbf{x}\rVert}}  =-\frac{1}{\cancel{\lVert \mathbf{x}\rVert^2}}\frac{\partial}{\partial \lVert\mathbf{x} \rVert} V\left(\lVert \mathbf{x}\rVert,x_s\right) \cancel{\langle \mathbf{x} ,\mathbf{x} \rangle} +\alpha\cos(\mathbf{x},\mathbf{u})\lVert \mathbf{u} \rVert
\end{equation}    
\hfill$\square$

\subsection{Proof of Lemma~\ref{lem:technical}}

If $\mathbf{x}_0=0$, since $\frac{\partial}{\partial r} V\left(0,x_s\right) = 0$ by Assumption~\ref{ass:V2}, $\mathbf{u} \neq 0$, and since the gradient system \eqref{eq:sys1} is continuous, there exists $t_0,\varepsilon \in \mathbb{R}_{>0}$ such that $\lVert \mathbf{x}(t_0) \rVert \geq \varepsilon$. Hence, the assumption $\mathbf{x}_0 \neq 0$ is made without loss of generality. It is easily noticed that if $x_{s,0} \geq 0$ in \eqref{eq:sys}, then $x_s(t) \geq 0$ for all $t\geq 0$, meaning the lower-bound~\eqref{eq:ineq_ass1} of Assumption~\ref{ass:V1} can be leveraged in \eqref{eq:2dreduced_dyn1} to obtain $\tau d\lVert \mathbf{x}\rVert/dt  \leq - a\lVert \mathbf{x}\rVert+ b+ \alpha\lVert \mathbf{u} \rVert$. Grönwall's lemma~\cite{Gronwall1919} then gives
\begin{equation}
    \lVert \mathbf{x}\rVert \leq  \lVert \mathbf{x}_0 \rVert e^{- a t/\tau}+\left(b+ \alpha\lVert \mathbf{u} \rVert\right)(1-e^{- at/\tau})/a
\end{equation}
It follows that for all $\mathbf{x}_0$ there exists $\overline{r}\in\mathbb{R}_{>0}$ upper-bounding $\lVert \mathbf{x} \rVert$. Using Grönwall's lemma again on the slow dynamics \eqref{eq:sys2} gives $x_s \in [0,\lVert x_{s,0}\rVert +g\left(\overline{r}\right)]$. The focus is now put on the dynamic of the scalar product between $\mathbf{x}$ and $\mathbf{u}$. Using the chain rule \eqref{eq:chainrule}, the following equalities hold:
\begin{equation}
    \tau\dot{\aoverbrace[L1R]{\langle \mathbf{x}, \mathbf{u} \rangle}} = \tau\langle \dot{\mathbf{x}}, \mathbf{u} \rangle  = -\frac{1}{\lVert \mathbf{x}\rVert} \frac{\partial}{\partial \lVert\mathbf{x} \rVert} V\left(\lVert \mathbf{x}\rVert,x_s\right)\langle \mathbf{x}, \mathbf{u} \rangle + \alpha\lVert \mathbf{u} \rVert^2
\end{equation}
If $\langle \mathbf{x}, \mathbf{u} \rangle \leq 0$, the Cauchy-Schwarz inequality and the lower-bound~\eqref{eq:ineq_ass1} of Assumption~\ref{ass:V1} yields $\tau\langle \dot{\mathbf{x}}, \mathbf{u} \rangle \geq \lVert \mathbf{u} \rVert \left( -b   + \alpha\lVert \mathbf{u} \rVert \right)$. Since $\lVert u \rVert > b/\alpha$, there exists $\mu \in \mathbb{R}_{>0}$ such that $\langle \dot{\mathbf{x}}, \mathbf{u} \rangle > \mu$, hence there exists a time $t_1$ above which $\langle \mathbf{x}, \mathbf{u} \rangle > 0$ for all $t\geq t_1$.
 Now, the upper-bound~\eqref{eq:ineq_ass2} of Assumption~\ref{ass:V1} yields $\tau\langle \dot{\mathbf{x}}, \mathbf{u} \rangle \geq -h(\overline{r})\langle \mathbf{x}, \mathbf{u} \rangle + \alpha \lVert \mathbf{u} \rVert^2$. It is easily demonstrated that:
\begin{equation}
     \tau\frac{d}{dt} \left(e^{ h(\overline{r})t/\tau}\langle \mathbf{x}, \mathbf{u} \rangle \right) \geq \alpha e^{h(\overline{r})t /\tau} \lVert \mathbf{u} \rVert^2
\end{equation}
Integrating both sides between $0$ and $t$ provides:
\begin{equation}
    \langle \mathbf{x}, \mathbf{u} \rangle \geq  \langle \mathbf{x}_0, \mathbf{u} \rangle e^{-h(\overline{r})t/\tau}+ \alpha\lVert \mathbf{u} \rVert^2(1-e^{-h(\overline{r})t/\tau})/h(\overline{r})
\end{equation}
Using Cauchy-Schwarz inequality hereabove shows that for all $t$ above a certain time $t_2$, there exists $\underline{r}\in\mathbb{R}_{>0}$ lower-bounding $\lVert \mathbf{x} \rVert$, thus concluding the proof.\hfill$\square$

  \subsection{Proof of Theorem 2}

For $\mathbf{x} \neq 0$ the following equality holds:
\begin{equation}
    \tau\dot{\aoverbrace[L1R]{ \mathbf{x}/\lVert \mathbf{x}\rVert }} = \frac{\tau}{ \lVert \mathbf{x} \rVert}\bigg(\dot{\mathbf{x}}  -\frac{\dot{\aoverbrace[L1R]{\lVert \mathbf{x}\rVert }}}{\lVert \mathbf{x} \rVert}  \mathbf{x}\bigg)
\end{equation}
substituting $\dot{\mathbf{x}}$ and $\dot{\aoverbrace[L1R]{\lVert \mathbf{x}\rVert }}$ respectively by \eqref{eq:sys1} and \eqref{eq:2dreduced_dyn1} provides:
\begin{equation}
   \tau\dot{\aoverbrace[L1R]{ \mathbf{x}/\lVert \mathbf{x}\rVert }} = - \cancel{\frac{1}{\lVert \mathbf{x} \rVert^2}\frac{\partial V}{\partial \lVert\mathbf{x}\rVert} \mathbf{x}}+\frac{\alpha}{\lVert \mathbf{x}\rVert}\mathbf{u} +\cancel{\frac{1}{\lVert \mathbf{x} \rVert^2}  \frac{\partial V}{\partial \lVert\mathbf{x} \rVert}\mathbf{x}}-\alpha \frac{\langle \mathbf{x},\mathbf{u} \rangle}{\lVert \mathbf{x} \rVert^3} \mathbf{x} 
\end{equation}
so the direction of $\mathbf{x}$ satisfies the differential equation
\begin{equation}
     \frac{\tau}{\alpha}\dot{\aoverbrace[L1R]{ \mathbf{x}/\lVert \mathbf{x}\rVert }} = \frac{1}{\lVert \mathbf{x} \rVert} \left(\mathbf{u}-\Big\langle\frac{\mathbf{x}}{\lVert \mathbf{x}\rVert}, \mathbf{u} \Big\rangle \frac{\mathbf{x}}{\lVert \mathbf{x}\rVert}  \right)
\end{equation}
Since $\mathbf{u}$ is constant
\begin{equation}
    \dot{\aoverbrace[L1R]{\cos(\mathbf{x},\mathbf{u}) }} = \Big\langle \dot{\aoverbrace[L1R]{ \frac{\mathbf{x}}{\lVert \mathbf{x}\rVert }}}, \frac{\mathbf{u}}{\lVert \mathbf{u} \rVert} \Big\rangle 
\end{equation}
which finally provides the Riccati differential equation:
\begin{equation}
    \frac{\tau}{\alpha}\dot{\aoverbrace[L1R]{\cos(\mathbf{x},\mathbf{u}) }}= \frac{1}{\lVert \mathbf{x} \rVert}\left( \frac{\lVert \mathbf{u} \rVert^{\cancel{2}}}{\cancel{\lVert \mathbf{u} \rVert}}-\lVert \mathbf{u} \rVert \cos^2(\mathbf{x},\mathbf{u}) \right)
\end{equation}
\hfill$\square$

\end{document}